\documentclass[aps,floatfix,showpacs,twocolumn]{revtex4-1}
\usepackage{amsmath}
\usepackage{amsfonts}
\usepackage{amssymb}
\usepackage{graphics,epsfig}
\usepackage{graphicx}
\begin{document}

\title{Some further notes on the Kruskal - Szekeres completion.}
\author{Kayll Lake}
\affiliation{Department of Physics, Queen's University, Kingston,
Ontario, Canada, K7L 3N6 }
\date{\today}
\begin{abstract}
In this pedagogical note, the Kruskal - Szekeres completion of the Schwarzschild spacetime is obtained explicitly in a few steps.
\end{abstract}
\maketitle
A discussion of the Kruskal \cite{Kruskal} - Szekeres \cite{Szekeres} completion of the Schwarzschild spacetime is a cornerstone of any modern introduction to general relativity. In a previous note \cite{lake}
I reviewed some history of these coordinates and showed how they can be obtained from Israel coordinates \cite{israel} and also from an integration of Einstein's equations. The purpose of the present note is to point out a very brief but direct and explicit derivation of the Kruskal - Szekeres completion. We use geometrical units and a signature of $+2$.

We start with the standard form of the Schwarzschild spacetime
\begin{equation}\label{line}
ds^2=-\left(1-\frac{2m}{r}\right)dt^2+\frac{dr^2}{1-\frac{2m}{r}}+r^2 d\Omega_2^2,
\end{equation}
where $d\Omega_2^2$ is the metric of a unit two-sphere and  $m$ is a constant $>0$. The radial null geodesic equations of (\ref{line}) are
\begin{equation}\label{geo}
   \frac{dr}{dt} = \pm \left(1-\frac{2m}{r}\right).
\end{equation}
The solutions to (\ref{geo}) can be given as
\begin{equation}\label{solns}
   r = 2m(\mathcal{L}(\Psi)+1)
\end{equation}
and
\begin{equation}\label{geo1}
    r = 2m(\mathcal{L}(\Phi)+1).
\end{equation}
Here $\mathcal{L}(x)$ is the Lambert W function \cite{lambert}, defined by $\mathcal{L}(x)\exp(\mathcal{L}(x)) = x$ \cite{lambert2},
\begin{equation}\label{Psi}
    \Psi^2 \equiv \left(\frac{u^2}{e}\right)^2\exp\left(\frac{t}{m}\right),
\end{equation}
\begin{equation}\label{Phi}
    \Phi^2 \equiv \left(\frac{v^2}{e}\right)^2\exp\left(-\frac{t}{m}\right),
\end{equation}
and $u$ and $v$ \textit{label} the radial  null geodesics. Now at any two - sphere $(u,v)$, where the branches of the radial null geodesics cross, $\Psi^2=\Phi^2$ and so from (\ref{Psi}) and (\ref{Phi}) we find
\begin{equation}\label{t}
    \exp\left(\frac{t}{m}\right)=\left(\frac{v}{u}\right)^2
\end{equation}
and so
\begin{equation}\label{Psi2}
    \Psi^2=\Phi^2=\left(\frac{uv}{e}\right)^2.
\end{equation}

\bigskip

Now in taking the appropriate roots to (\ref{t}) and (\ref{Psi2}) we must orientate the $u-v$ axes. Noting that $\mathcal{L}(-1/e)=-1$ and $\mathcal{L}(0)=0$ we take $uv>0$ over the range $0<r<2m$ and so $ \Psi=\Phi=-\frac{uv}{e}$. We have arrived at \cite{df}
\begin{equation}\label{ks2}
d\tilde{s}^2=\frac{-(4m)^2}{(1+\mathcal{L})\exp(1+\mathcal{L})}dudv+(2m)^2(1+\mathcal{L})^2d\Omega^2_{2},
\end{equation}
where we have now written $\mathcal{L}\equiv\mathcal{L}(-uv/e)$ and take $\exp(t/2m)=v/u$ for $uv>0$ and $\exp(t/2m)=-v/u$ for $uv<0$.  This is the null form of the Kruskal - Szekeres completion.

We convert our results into standard notation in the Appendix.
\begin{acknowledgments}
I thank Bill Unhru for useful criticism of an earlier draft. This work was supported by a grant from the Natural Sciences and
Engineering Research Council of Canada. Portions of this work were
made possible by use of \textit{GRTensorII} \cite{grt}.
\end{acknowledgments}

\section*{Appendix}
Following, for example \cite{mtw}, define
\begin{equation}\label{UVdef}
    V \equiv \frac{v+u}{2},\;\;\; U \equiv \frac{v-u}{2}
\end{equation}
so that
\begin{equation}\label{uv}
    v = V+U, \;\;\; u=V-U.
\end{equation}
Since
\begin{equation}\label{rr}\nonumber
  r = 2m(1+\mathcal{L}(-\frac{uv}{e})), \;\;\ \mathcal{L}\exp(\mathcal{L})=-\frac{uv}{e}
\end{equation}
we have
\begin{equation}\label{usual}
    \left( \frac{r}{2m}-1 \right)\exp\left(\frac{r}{2m}\right) = U^2 - V^2,
\end{equation}
the usual implicit relation for $r$. We now have \cite{df}
\begin{equation}\label{usualline}
    ds^2 = -\frac{32m^3}{r}\exp\left(-\frac{r}{2m}\right)\left(dV^2-dU^2 \right) +r^2 d\Omega_2^{2},
\end{equation}
the usual form of the Kruskal - Szekeres metric.

Finally, for $uv>0$ we have $v/u=\exp(t/2m)$ and so
\begin{equation}\label{UV}
    \frac{U}{V}  = \tanh \left( \frac{t}{4m} \right), \;\;\; V^2 > U^2
\end{equation}
and  for $uv<0$ we have $v/u=-\exp(t/2m)$ and so
\begin{equation}\label{VU}
    \frac{V}{U}  = \tanh \left( \frac{t}{4m} \right), \;\;\; V^2 < U^2.
\end{equation}
The usual procedure is to introduce (\ref{usual}), (\ref{UV}) and (\ref{VU}) as coordinate transformations to change (\ref{line}) to (\ref{usualline}). The purpose of this note is to point out that these seemingly mysterious coordinate transformations need never be introduced.


\begin{thebibliography}{}\label{sec:TeXbooks}
\bibitem{Kruskal} M. Kruskal, Phys. Rev., \textbf{119}, 1743 (1960).
\bibitem{Szekeres}G. Szekeres, Gen. Rel. Grav., \textbf{34}, 2001 (2002) (Reprinted from Publicationes Mathematicae Debrecen \textbf{7}, 285 (1960)).
\bibitem{lake} K. Lake, Class. Quantum Grav., \textbf{27}, 097001 (2010), [arXiv:gr-qc/1002.3600].
\bibitem{israel} W. Israel, Phys. Rev., \textbf{143}, 1016 (1966).
\bibitem{lambert}See, for example, R. M. Corless,
G. H. Gonnet, D. E. G. Hare, D. J. Jeffrey, and D. E. Knuth,
Advances in Computational Mathematics \textbf{5}, 329 (1996).
\bibitem{lambert2}Note that $x \geq -1/e$. We require $\mathcal{L}(x)\geq-1$ so that $\mathcal{L}(x)$ is single valued.
\bibitem{df}Note that $d\mathcal{L}/dz=\mathcal{L}/z(1+\mathcal{L})$.
\bibitem{mtw}C. W. Misner, K. S. Thorne and J. A. Wheeler, \textit{Gravitation} (W. H. Freeman, San Francisco, 1973).

\bibitem{grt}This package runs within Maple. The GRTensorII software and documentation is
distributed freely from the address \textit{http://grtensor.org}
\end{thebibliography}
\end{document}